\documentstyle[12pt,epsf]{article}

\newlength{\pubnumber} 

\catcode`\@=11
\@addtoreset{equation}{section}

\def\section{\@startsection{section}{1}{\z@}{3.5ex plus 1ex minus .2ex}
 {2.3ex plus .2ex}{\large\bf}}
\def\subsection{\@startsection{subsection}{2}{\z@}{2.3ex plus .2ex}
 {2.3ex plus .2ex}{\bf}}

\def\beq{\begin{equation}}
\def\eeq{\end{equation}}
\def\beqn{\begin{eqnarray}}
\def\eeqn{\end{eqnarray}}

\def\half{{\textstyle{1\over 2}}}
\def\third{{\textstyle {1\over3}}}

\def\twothird{{\textstyle {2\over3}}}

\def\I{{\bf I}}
\def\mbf{\mathbf}
\def\bone{{\mathbf 1}}

\def\bo{{\mathbf 0}}

\def\bS{{\mathbf S}}
\def\bV{{\mathbf V}}

\def\inbar{\,\vrule height1.5ex width.4pt depth0pt}
\def\IT{\relax\hbox{$\inbar\kern-.3em{\rm T}$}}
\def\IS{\relax\hbox{$\inbar\kern-.3em{\rm S}$}}
\def\IC{\relax\hbox{$\inbar\kern-.3em{\rm C}$}}
\def\IQ{\relax\hbox{$\inbar\kern-.3em{\rm Q}$}}
\def\IR{\relax{\rm I\kern-.18em R}}
 \font\cmss=cmss10 \font\cmsss=cmss10 at 7pt

\def\IZ{\relax\ifmmode\mathchoice
 {\hbox{\cmss Z\kern-.4em Z}}{\hbox{\cmss Z\kern-.4em Z}}
 {\lower.9pt\hbox{\cmsss Z\kern-.4em Z}}
 {\lower1.2pt\hbox{\cmsss Z\kern-.4em Z}}\else{\cmss Z\kern-.4em Z}\fi}

\def\Io{\relax\ifmmode\mathchoice
 {\hbox{\cmss 1\kern-.4em 1}}{\hbox{\cmss 1\kern-.4em 1}}
 {\lower.9pt\hbox{\cmsss 1\kern-.4em 1}}
{\lower1.2pt\hbox{\cmsss 1\kern-.4em 1}}\else{\cmss 1\kern-.4em 1}\fi}

\begin{document}
\begin{titlepage}
\setcounter{page}{1}
\rightline{BU-HEPP-08-16}
\rightline{CASPER-08-06}

\vspace{.06in}
\begin{center}
{\Large \bf A Non-Standard String Embedding of $E_8$}
\vspace{.12in}

{\large 
Richard Obousy,\footnote{Richard\_K\_Obousy@baylor.edu}
Matthew Robinson,\footnote{M\_Robinson@baylor.edu} and
Gerald B. Cleaver\footnote{Gerald\_Cleaver@baylor.edu}}
\\
\vspace{.12in}
{\it Center for Astrophysics, Space Physics \& Engineering Research\\
            Department of Physics, Baylor University,
            Waco, TX 76798-7316\\}
\vspace{.06in}
\end{center}

\begin{abstract}
An algorithm to systematically and efficiently generate free fermionic heterotic string models was recently introduced
\cite{mr}. This algorithm has been adopted by the Free Fermionic Model Construction (FFMC) program at Baylor University. 
As its first application, the algorithm is being applied to  
systematically generate the {\it complete} set of free fermionic heterotic string models with untwisted left-moving (worldsheet supersymmetric) sectors, up to continually advancing Layer and Order. Statistical analysis of this study will be reported in the near future.
However, in a series of separate notes we will be reporting some of the more interesting models that appear along the way. 
In this, our first such note, we reveal a different string embedding of $E_8$ than is standard.
That is, rather than realize $E_8$ via an $SO(16)$ embedding, $\mbf{248} = \mbf{120} + \mbf{128}$, we realize it via an $SU(9)$ embedding, $\mbf{248} = \mbf{80} + \mbf{84} + \overline{\mbf{84}}$. This is obtained in a Layer 1, Order 6 model for which modular invariance itself dictates a gravitino sector accompany the gauge sector.
\end{abstract}
\end{titlepage}
\setcounter{footnote}{0}

\section{Systematic Investigation of Free Fermionic Heterotic String Models}

An algorithm \cite{mr} to systematically and efficiently generate free fermionic \cite{fff1,fff2} heterotic models was recently introduced. As a first application, we have initiated an indepth study of the statistics of the gauge groups in free fermionic heterotic strings with only untwisted left-moving (worldsheet supersymmetric) sectors. Our approach enables a {\it complete} study of all gauge group models 
to be generated and analyzed with extreme efficiency, up to continually increasing Layers (the number of gauge basis vectors) and Orders (the lowest positive integer $N$ that transforms, by multiplication, each basis vector back into the untwisted sector mod(2)). 
In this initial study the models have either ${\cal{N}}=4$ or ${\cal{N}}=0$ spacetime SUSY, depending on whether the gravitino sector is or is not present, respectively.
The primary goal of our research is to systematically improve the understanding of the statistical properties and characteristics of free fermionic heterotic models, a process that is underway by a collection of research groups \cite{af1,kd}. 
However, as particularly interesting models appear in the course of our program, we will separately report on such models. The first of these models appears at Layer 1, Order 6 and requires a graviton sector. The intersting feature of this model is that it provides an alternative embedding of $E_8$, based not on the $E_8$ maximal subgroup $SO(16)$, but on $E_8$'s alternate maximal subgroup $SU(9)$.

\section{Review of $E_8$ String Models in 4 and 10 Dimension}

The $SO(16)$ realization of $E_8$ is well known: We start with the uncompactified $D=10$, ${\cal{N}}=1$ SUSY $SO(32)$ heterotic string in light-cone gauge. 
Free fermion construction generates this model from two basis boundary vectors: the ever-present all-periodic vector, $\bone$, and the supersymmetry generating vector $\bS$ \cite{fff1}:
\beqn
\bone &=& [(1)^{8}|| (1)^{32}]\label{aps10}\\
\bS   &=& [(1)^{8}|| (0)^{32}]\label{susy10}.
\eeqn
The $\mbf 496$ (adjoint) rep of $SO(32)$ is produced by the untwisted boundary vector $\bo = \bone+\bone$,
\beqn
\bo &=& [(0)^{8}|| (0)^{32}]\label{unt10}.
\eeqn

To transform the uncompactified $D=10$, ${\cal{N}}=1$ SUSY $SO(32)$ heterotic model into the $D=10$, ${\cal{N}}=1$ SUSY $E_8\otimes E_8$ model,
all that is required is the additional twisted basis boundary vector \cite{fff1},
\beqn
\I^O =  [(0)^{8}|| (1)^{16} (0)^{16}].
\label{tsistO10}
\eeqn
The GSO projection of $\I^O$ onto $\bo$ reduces the untwisted sector gauge group to $SO(16)_O\otimes SO(16)_H$
by reducing its massless gauge states to the adjoint reps $\mbf{120}_O\otimes \mbf{1}$ + $\mbf{1}\otimes \mbf{120}_H$.
The GSO projection of $\I^O$ (or of $\bone$) on $\I^O$ results in a $\mbf{128}_O\otimes 1$ massless spinor rep of definite chirality.
Further, the GSO projection of $\bone$ onto 
\beqn
\I^{H} = \I^{O} + \bone + \bS =  [(0)^{8}|| (0)^{16} (1)^{16}],
\label{tsistH10}
\eeqn
produces a massless spinor rep $\mbf{1}\otimes \mbf{128}_H$ of $SO(16)_H$ with matching chirality.

Thus, the boundary sectors $\bo$ and $\I^O$ produce the 
 ${\mbf 248}$ (adjoint) of an observable $E_8$ via the $SO(16)$ embedding 
\beqn
\mbf{248} = \mbf{120} + \mbf{128},
\label{so16Oemb}
\eeqn
while the boundary sectors $\bo$ and $\I^H$ produce the same for a hidden sector $E_8^H$

When the $E_8\otimes E_8$ model is compactified to four dimensions, without any twist applied to the compact dimensions, 
the basis vectors become,
\beqn
\bone &=& [(1)^{2}, (1,1,1)^{6}|| (1)^{44}]\label{aps4}\\
\bS   &=& [(1)^{2}, (1,0,0)^{6}|| (0)^{44}]\label{susy4}\\
\I^O  &=& [(0)^{2}, (0,0,0)^{6}|| (1)^{16} (0)^{28}].\label{tsist4}
\eeqn
Because
\beqn
\I^H  = \I^O + \bone +\bS &=& [(0)^{2}, (0,1,1)^{6} || (0)^{16} (1)^{28}],
\label{tsihst4}
\eeqn
is no longer a massless sector, the gauge group is $E_8^O\otimes SO(22)$ (with ${\cal{N}}=4$ SUSY).
An additional massless twisted sector,
\beqn
\I^{H'}  &=& [(0)^{0}, (0,0,0)^{6|}| (0)^{16}, (1)^{16}, (0)^{6}],
\label{tsihpst4}
\eeqn
is required to reclaim the second $E_8$.\footnote{In this note we we do not discuss the gauge group of the left-moving sector, since it belongs 
to the $N=4$ gravity multiplet and disappears for $N<2$.}

\section{$E_8$ from $SU(9)$} 

Our systematic research of free fermionic gauge models, revealed at Layer 1, Order 3 (more precisely Layer 1, Order 6 = Order(2) x Order(3))
as explained below) an intersting alternative realization of $E_8$ 
The simplest possible massless gauge sector for Order 3 is
\beqn
\I^{3}  &=& [(0)^{0}, (0,0,0)^{6|}| (\twothird)^{18}, (0)^{26}].
\label{tsi34}
\eeqn
The non-integer values in $\I^{3}$ produce a GSO projection on the untwisted sector that breaks $SO(44)$ down to
$SU(9)\otimes U(1)\otimes SO(26)$. The charges of the $SU(9)$ non-zero roots are of the form 
$\pm(1_i,-1_j)$ for $i$ and $j\ne i$ denoting one of the first 9 right-moving complex fermion. Combined with the 
zero roots of the Cartan Subalgrabra, these form the $\mbf 80$ (adjoint) rep of $SU(9)$
The $U(1)= \sum{i=1}^{9} U(1)_i$ charge is ${\rm Tr}\, Q_i$. The $SO(26)$ generators have the standard charges of
$\pm(1_r,\pm 1_s)$ with $r$ and $s\ne r$ denoting one of the last 13 right-moving complex fermion.  

However, two of the the modular invariance requirements for basis vectors $\bV_i$ and $\bV_j$ \cite{fff1}, specifically
\beqn
N_{i,j} \bV_i\cdot \bV_j &=& 0\,\, ({\rm mod}\, 4),\,\, {\rm and} \label{mi1}\\
N_i \bV_i\cdot \bV_i     &=& 0\,\, ({\rm mod}\, 8), \label{mi2}
\eeqn
necessitate that $\I^3$ be expressed as a spacetime fermion, rather than spacetime boson. That is, the required
basis boundary vector to produce a gauge factor of $SU(9)$ in the untwisted sector in like manner to (\ref{tsi34}) is
\beqn
\I^{6}  &=& [(1)^{0}, (1,0,0)^{6|}| (\twothird)^{18}, (0)^{26}].
\label{tsi64}
\eeqn 
As an Order 6 = Order 2 x Order 3, basis boundary vector,
(\ref{tsi64}) satisfies (\ref{mi1},\ref{mi2}). $2\I^{6} = \I^{3}$ is then a massless gauge sector, as is
$4\I^{6}= -\I^{3}$. Note also that $3\I^6$ is the gravitino sector $S$. Hence $\bS$ need not, and cannot, be a 
separate basis vector.

The GSO projections of $\bone$ and $\I^{6}$ on $I^{3}$ and $-I^{3}$ yield massless gauge states from two sets of charges. Charges in the
first set have the form 
\beqn
\pm(-\twothird_{i_1},-\twothird_{i_2},-\twothird_{i_3},\third_{i_4},\third_{i_5},\third_{i_6},\third_{i_7},\third_{i_8},\third_{i_9}),
\label{set1}
\eeqn
with all subscripts different and each denoting one of the first 9 complex fermions.  
States in $I^{3}$ and $-I^{3}$ vary by their overall charge sign and form the $\mbf{84}$ and $\overline{\mbf{84}}$ reps of $SU(9)$, respectively
Thus, together the sectors $\bo$, $I^{3}$, and $-I^{3}$ contain the $\mbf{80}$, $\mbf{84}$ and $\overline{\mbf{84}}$ reps of $SU(9)$, from which
$\mbf{248} = \mbf{80} + \mbf{84} + \overline{\mbf{84}}$ emerges. Thus, here $E_8$ is obtained from its second maximal subgroup $SU(9)$. 

The second set of charges are of the form,
\beqn
\pm(\third,\third,\third,\third,\third,\third,\third,\third,\third, \pm 1_r),\label{soe}
\label{set2}
\eeqn
with $1_r$ denoting a unit charge of one of the 13 complex fermions generating the $SO(26)$ Cartan subalgebra. 
Hence, the charges in this set are orthogonal to $E_8$, but have non-zero dot products with $U(1)= \sum{i=1}^{9} U(1)_i$ charged states, and unit dot products with the $SO(26)$ generators. Thus, this second set of states enhance $SO(26)$ to $SO(28)$.
The complete gauge group is thus $E_8\otimes SO(28)$. Since the gravitino sector is a multiple of $I^{6}$, the model has inherent ${\cal{N}}=4$ SUSY.

The whole process can be followed again with the addition of another basis boundary vector $\I^{6H}$ isomorphic with $\I^{6}$, but that has no non-zero right-moving charges in common with $\I^{6}$:
\beqn
\I^{6H}  &=& [(1)^{0}, (1,0,0)^{6|}| (0)^{18}, (\twothird)^{18}, (0)^{8}].
\label{tsi64h}
\eeqn 
$\I^{6H}$ will produce a second $E_8$ from a parallel $SU(9)$ embedding. The $SO(8)$ of the untwisted sector would be enhanced by both of the
$U(1)$'s associated with the two $SU(9)$'s to $SO(12)$, giving a standard $E_8\otimes E_8 \otimes SO(12)$ model, but with an $SU(9)\times SU(9)$ embedding for $E_8\otimes E_8$.

Heterotic models have an $SO(44)$ rotational redundancy in their charge expressions (which we are taking into account in our 
statistical analysis). In terms of solely the gauge sectors, our $E_8$ embedding from $SU(9)$ can be understood as a specific
$SO(18)\in SO(44)$ rotation of the initial charge lattice. In the $SO(16)$ basis, a set of simple roots for $E_8$ are
\beqn
E_1 &=& (+1,-1, 0, 0, 0, 0, 0, 0)\label{eq1}\\
E_2 &=& ( 0,+1,-1, 0, 0, 0, 0, 0)\label{eq2}\\
E_3 &=& ( 0, 0,+1,-1, 0, 0, 0, 0)\label{eq3}\\
E_4 &=& ( 0, 0, 0,+1,-1, 0, 0, 0)\label{eq4}\\
E_5 &=& ( 0, 0, 0, 0,+1,-1, 0, 0)\label{eq5}\\
E_6 &=& ( 0, 0, 0, 0, 0,+1,-1, 0)\label{eq6}\\
E_7 &=& ( 0, 0, 0, 0, 0,+1,+1, 0)\label{eq7}\\
E_8 &=& (-\half,-\half,-\half,-\half,-\half,-\half,-\half,-\half),\label{eq8}
\eeqn
where we choose a positive chirality $\mbf{128}$ spinor.
For an $SO(18)$ rotation we need 9 charge states, so we we will add
an zero charge onto the $E_8$ charges and include a
$U(1)$ generator with defining charge generator
\beqn
U_9 = ( 0, 0, 0, 0, 0, 0, 0, 0, 1).\label{eq9}
\eeqn

Alternately, a simple set of roots for the $SU(9)$ basis is
\beqn
E^{'}_1 &=&(+1,-1, 0, 0, 0, 0, 0, 0, 0)\label{eq1p}\\
E^{'}_2 &=&( 0,+1,-1, 0, 0, 0, 0, 0, 0)\label{eq2p}\\
E^{'}_3 &=&( 0, 0,+1,-1, 0, 0, 0, 0, 0)\label{eq3p}\\
E^{'}_4 &=&( 0, 0, 0,+1,-1, 0, 0, 0, 0)\label{eq4p}\\
E^{'}_5 &=&( 0, 0, 0, 0,+1,-1, 0, 0, 0)\label{eq5p}\\
E^{'}_6 &=&( 0, 0, 0, 0, 0,+1,-1, 0, 0)\label{eq6p}\\
E^{'}_7 &=&( 0, 0, 0, 0, 0, 0,+1, 0,-1)\label{eq7p}\\
E^{'}_8 &=&(-\third,-\third,-\third,-\third,-\third, \twothird, \twothird,-\third,\twothird).\label{eq8p}
\eeqn
In the $SU(9$ basis, there is also an additonal $U(1)$ of the form
\beqn
U^{'}_9 =(\third,\third,\third,\third,\third,\third,\third,\third,\third).\label{eq9p}
\eeqn

The $SO(16)$ embedding of $E_8$ can be transformed into the $SU(9)$ embedding of $E_8$ via a Weyl rotation
that yields 
\beqn
E^{'}_7 &=& \half(E_7 - E_6) - U_9 \label{eq7pw}\\
E^{'}_8 &=& \twothird(E_8 + U_9) \label{eq8pw}\\
U^{'}_9 &=& \twothird(-E_8 + \half U_9). \label{eq9pw}
\eeqn
Note also that the rotation between these $E_8$ embeddings can be expressed in terms of partition function equivalences 
involving Theta-function product identities \cite{mumford}.

\section{Summary}

In this note we presented an alternative embedding for $E_8$, involving not its maximal subgroup $SO(16)$, rather its alternate
maximal subgroup $SU(9)$. Instead of the $\mbf{248}$ (adjount) rep of $E_8$ 
generated as $\mbf{248} = \mbf{120} + \mbf{128}$ of $SO(16)$, we constructed a $D=4$ model in which it is generated as 
$\mbf{248} = \mbf{80} + \mbf{84} + \overline{\mbf{84}}$ of $SU(9)$. Interestingly, we found that in this model that modular invariance requires the basis boundary vector responsibble for the pair of massless gauge sectors that yields the $\mbf{84} + \overline{\mbf{84}}$ reps to also
produce the gravitino-producing sector. The model starts out with ${\cal{N}}=4$ SUSY. Thus, this alternate $E_8$ embedding cannot occur in a model without either broken or unbroken SUSY (i.e., a model that lacks a gaugino sector).

\section*{Acknowledgments}

Research funding leading to this manuscript was partially provided by Baylor URC grant 0301533BP.
\vfill
\newpage

\def\AIP#1#2#3{{\it AIP Conf.\ Proc.}\/{\bf #1} (#2) #3}
\def\AP#1#2#3{{\it Ann.\ Phys.}\/ {\bf#1} (#2) #3}
\def\IJMP#1#2#3{{\it Int.\ J.\ Mod.\ Phys.}\/ {\bf A#1} (#2) #3}
\def\IJMPA#1#2#3{{\it Int.\ J.\ Mod.\ Phys.}\/ {\bf A#1} (#2) #3}
\def\JHEP#1#2#3{{\it JHEP}\/ {\bf #1} (#2) #3}
\def\MODA#1#2#3{{\it Mod.\ Phys.\ Lett.}\/ {\bf A#1} (#2) #3}
\def\MPLA#1#2#3{{\it Mod.\ Phys.\ Lett.}\/ {\bf A#1} (#2) #3}
\def\NJP#1#2#3{{\it New\ J.\ Phys.}\/ {\bf #1} (#2) #3}
\def\nuvc#1#2#3{{\it Nuovo Cimento}\/ {\bf #1A} (#2) #3}
\def\NPB#1#2#3{{\it Nucl.\ Phys.}\/ {\bf B#1} (#2) #3}
\def\NPBPS#1#2#3{{\it Nucl.\ Phys.}\/ {{\bf B} (Proc. Suppl.) {\bf #1}} (#2)#3}
\def\PLB#1#2#3{{\it Phys.\ Lett.}\/ {\bf B#1} (#2) #3}
\def\PRD#1#2#3{{\it Phys.\ Rev.}\/ {\bf D#1} (#2) #3}
\def\PRL#1#2#3{{\it Phys.\ Rev.\ Lett.}\/ {\bf #1} (#2) #3}
\def\PRT#1#2#3{{\it Phys.\ Rep.}\/ {\bf#1} (#2) #3}
\def\PTP#1#2#3{{\it Prog.\ Theo.\ Phys.}\/ {\bf#1} (#2) #3}
\def\RPP#1#2#3{{\it Rept.\ Prog.\ Phys.}\/ {\bf #1} (#2) #3}
\def\etal{{\it et al\/}}


\end{document}